# Collaborative Applications over Peer-to-Peer Systems – Challenges and Solutions

H. M. N. Dilum Bandara and Anura P. Jayasumana

Colorado State University, Fort Collins, CO 80523.
dilumb@engr.colostate.edu, Anura.Jayasumana@colostate.edu

## Abstract

Emerging collaborative Peer-to-Peer (P2P) systems require discovery and utilization of diverse, multi-attribute, distributed, and dynamic groups of resources to achieve greater tasks beyond conventional file and processor cycle sharing. Collaborations involving application specific resources and dynamic quality of service goals are stressing current P2P architectures. Salient features and desirable characteristics of collaborative P2P systems are highlighted. Resource advertising, selecting, matching, and binding, the critical phases in these systems, and their associated challenges are reviewed using examples from distributed collaborative adaptive sensing systems, cloud computing, and mobile social networks. State-of-the-art resource discovery/aggregation solutions are compared with respect to their architecture, lookup overhead, load balancing, etc., to determine their ability to meet the goals and challenges of each critical phase. Incentives, trust, privacy, and security issues are also discussed, as they will ultimately determine the success of a collaborative P2P system. Open issues and research opportunities that are essential to achieve the true potential of collaborative P2P systems are discussed.

# 1 Introduction

Resource-rich computing devices, decreasing communication cost, and Web 2.0 technologies are fundamentally changing the way we communicate, learn, socialize, and collaborate. With these changes, we envision Peer-to-Peer (P2P) systems that play an even greater role in collaborative applications. P2P computing fits naturally to this new era of user-driven, distributed applications utilizing resource-rich edge devices. There is thus a tremendous opportunity to create value by combining societal trends with P2P systems. Peer collaboration has expanded beyond its conventional applications wherein files or processor cycles are shared by peers to perform similar tasks. Emerging collaborative P2P systems look for diverse peers that could bring in unique capabilities to a community, thereby empowering it to engage in greater tasks beyond that can be accomplished by individual peers, yet are beneficial to all the peers. This is similar to a team that thrives due to the diversity of members' expertise. A *collaborative P2P system* is a P2P system that aggregates a group(s) of diverse resources (e.g., hardware, software, services, and data) to accomplish a greater task. Such collaborations involving diverse and application specific resources and dynamic Quality of Service (QoS) goals stress the current P2P architectures.

Collaborative P2P systems are applicable in a wide variety of contexts such as Distributed Collaborative Adaptive Sensing (DCAS) [1], grid/cloud computing, opportunistic computing [2], Internet of Things, social networks, and emergency management. We consider four representative applications to illustrate the salient features and characteristics of collaborative P2P systems. First is Collaborative Adaptive Sensing of the Atmosphere (CASA) [1, 3], an emerging DCAS system based on a dense network of weather radars that collaborate in real time to detect hazardous atmospheric conditions such as tornados and severe storms. CASA also employs many small sensors such as pressure sensors and rain gauges to further enhance the detectability and prediction accuracy of weather events. Collaborative P2P data fusion provides an attractive implementation choice for CASA real-time radar data fusion [3], wherein data are constantly being generated, processed, and pushed and pulled among groups of heterogeneous radars, sensors, storage, and processing nodes. The group of radars, sensors, processing, and storage elements involved in tracking a particular weather event may continue to change as the weather event migrates in both time and space. Thus, new groups of resources may have to be aggregated and current resources may be released as and when needed. Moreover, certain rare but severe weather events require specific meteorological algorithms (e.g., signal processing and prediction) and more computing, storage, and bandwidth resources to track and forecast/nowcast[*] about the behavior of those events. It is neither feasible nor economical to provision resources for such rare peak demands everywhere in the CASA system. Instead, a collaborative P2P system can exploit the temporal and spatial diversity of weather events to aggregate underutilized resources from anywhere in the system as far as desired performance and QoS goals are satisfied. On the other end of the spectrum, we are also seeing the emergence of crowd-sourced, community-based weather monitoring systems [4] that aggregate armature weather stations and community-based computing resources to provide local and national weather forecasts. The second application is cloud computing, which is transforming the way we host and run applications because of its rapid scalability and pay-as-you-go economic model [5]. Open cloud initiatives are pressing for interoperability among multiple cloud providers and sites of the same provider. Alternatively, community cloud computing [6], based on underutilized computing resources in homes or businesses for example, targets issues such as centralized data, privacy, proprietary applications, and cascading failures in modern clouds. Certain applications also benefit from a mixture of dedicated and voluntary cloud resources [7-8]. A collaborative P2P system is the core of such a multi-site or community-cloud system that interconnects dedicated/voluntary resources while dealing with rapid scalability and resource fluctuations. We refer to such systems as *P2P clouds*. Third application, Global Environment for Network Innovations (GENI) [9], is a collaborative and exploratory platform for discoveries and innovation. GENI allows users to aggregate diverse resources (e.g., processing, networks, sensors, actuators, and software) from multiple administrative domains for a common task. A collaborative P2P system is a natural fit for GENI because of its distributed, dynamic, heterogeneous, and collaborative nature. Fourth, the value of social networks can be enhanced by allowing users to share diverse resources available in their mobile devices [2]. For example, a person with a basic mobile phone could connect to a friend's smart phone with GPS capability to locate a nearby ATM. In another example, a group sharing their holiday experiences in a coffee shop could use one member's projection phone to show pictures from others' mobiles/tablets or stream videos from their home servers. Moreover, in large social gatherings such as carnivals, sports events, or political rallies, users' mobile devices can be used to share hot deals, comments, videos, or vote for a certain resolution without relying on a network infrastructure. Such applications are already emerging under the domain of opportunistic networking and computing [2]. These applications, hereafter referred to as *mobile social networks*, also benefit from collaborative mobile P2P technology. Further, envision an agglomeration of many collaborative P2P systems into a single unified P2P framework wherein peers contribute and utilize diverse resources for both altruistic and commercial purposes. For example, a cloud provider could contribute its processor cycles to the P2P community hoping to gain monetary benefits whenever possible, and during periods of lower demand it could provide similar or degraded services to gain nonmonetary benefits (e.g., to demonstrate its high availability or gain reputation). Alternatively, an application provider that accesses free/unreliable resources for its regular operations could tap into dedicated/reliable resources during periods of high demand [7]. The framework could also enable resource-rich home users to earn virtual currency[†] or points for their contributions that they can later use to access other services offered within the system [10-11]. Such a framework also enables a level playing field for both small and large-scale contributors.

---

[*] Nowcast refers to a 5-30 minute short-term forecast of an active weather event.

CASA, P2P clouds, GENI, mobile social networks, and aggregated P2P systems depend on some form of resource collaboration. These systems share a variety of *resources* such as processor cycles, storage capacity, network bandwidth, sensors/actuators, special hardware, middleware, scientific algorithms, application software, services (e.g., web services and spawning nodes in a cloud), and data to not only consume a variety of contents but also to generate, modify, and manage those contents. Moreover, these resources are characterized by multiple static and dynamic attributes. For example, CPU speed, free CPU capacity, memory, bandwidth, operating system, and a list of installed applications/middleware and their versions may characterize a processing node. These multi-attribute resources need to be combined in a timely manner to meet the performance and QoS requirements of collaborative P2P applications. Yet, it is nontrivial to discover, aggregate, and utilize heterogeneous and dynamic resources that are distributed.

---

† Electronic money that is not a tangible commodity and are typically not contractually backed by tangible assets nor by legal laws.

This paper attempts to formalize the P2P resource collaboration problem and reviews the current solution space. Key phases and desirable characteristics of collaborative P2P systems are identified first. We then evaluate the (in)ability of existing solutions' to support those key phases and desired characteristics. Incentives, trust, privacy, and security related issues and existing solutions are also discussed. Finally, a detailed discussion on open issues and research opportunities is provided. Key phases and desirable characteristics of resource collaboration are introduced in Section 2. Section 3 discusses how different solutions for structured and unstructured P2P systems are designed to provide the functionality required by each phase. Section 4 presents incentives, trust, privacy, and security issues and solutions proposed to address those. Research opportunities and challenges in resource collaboration over P2P systems are discussed in Section 5. Concluding remarks are presented in Section 6.

## 2 Phases in Resource Collaboration

We identify seven phases of resource collaboration in P2P systems as shown in Figure 1. These separate phases are identified for the sake of conceptual understanding while an actual implementation may combine multiple phases together. Specific systems may skip some of them depending on the application requirements. However, complexities, implementation details, and available related work on each of these phases significantly vary. The seven phases are as follows:

1. *Advertise* – Each node advertises its resources and their capabilities using one or more *Resource Specifications* (RSs). Taxonomy of RSs is given in Fig 2(a). A resource is characterized by a set of attributes $A$ and a typical RS is as follows:

$$RS = (a_1 = v_1, a_2 = v_2, ..., a_i = v_i)$$

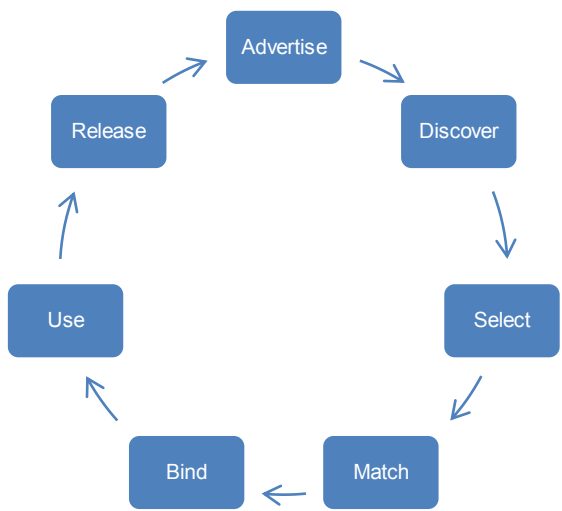

**Figure 1.** Phases in resource collaboration.

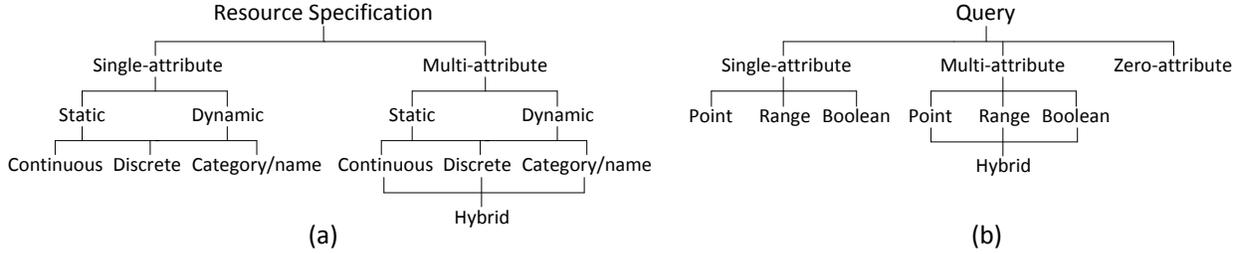

**Figure 2.** Taxonomy of (a) Resource specifications and (b) Queries.

Each attribute $a_i \in A$ has a value $v_i \in D_i$ that belongs to a given domain $D_i$. $D_i$ is typically bounded and may represent a continuous/discrete value or a category/name. For example, free CPU is continuous, number of CPU cores is discrete, Doppler radar is a category, and a file is represented by a name. In some cases, $v_i$ may specify a range or list of values, e.g., a time range or list of software libraries. Attribute values are further classified as static (e.g., CPU speed, operating system, and Doppler radar) and dynamic (e.g., CPU and memory free). RSs are also classified based on the number of attributes. RSs are referred to as single-attribute when the cardinality of set $A$, $|A| = 1$ and they are referred to as multi-attribute when $|A| > 1$. Some resources are advertised using a single-attribute that provides an abstract representation of the resource. For example, cloud-computing nodes are typically described as *high CPU*, *high memory*, and *cluster* instances [12]:

$$RS = (Type = "High\ CPU")$$

RSs may also specify constraints on how those resources can be used. For example, a computing node may indicate when it is available, to what extent the CPU can be utilized, who can use it, and how many concurrent licenses are available. Hence, a detailed hybrid, multi-attribute RS may look like:

$$RS = \begin{pmatrix} CPUSpeed = 2.0\ GHz,\ Architecture = \times 86,\ CPUFree = 53\%,\ MemoryFree = 1071\ MB, \\ OS = Linux\_2.6.31,\ Available = 12:00\ am - 6:00\ am,\ CPUUtilization \leq 60\%, \\ UseBy = "Friends",\ Licenses = 2 \end{pmatrix}$$

In practice, a RS is advertised as a tuple, set of (attribute, value) pairs, vector [13], or as a detailed XML specification [14]. A node may either hold on to its RSs hoping others will discover its resources by sending probe messages, or explicitly send/advertise its RSs to a centralized database, Distributed Hash Tables (DHTs), or to a set of random nodes. Resources that are characterized by dynamic attributes are typically re-advertised when the attribute values change significantly.

2. *Discover* – Nodes may send probing messages to proactively discover and build a local repository of useful RSs, particularly if specifications are unadvertised. Such a collection of RSs can speed up the query resolution and may be used to keep track of inter-resource relationships such as latency, bandwidth, and trust.

3. *Select* – Select a group(s) of resources that satisfies the given application requirements. Application requirements are typically specified using queries that list one or more attributes and range of attribute values. The taxonomy of queries is given in Fig. 2(b). A typical range query may be defined as follows:

$$Q = (Resources = m, a_1 \in [l_1, u_1], a_2 \in [l_2, u_2], \ldots, a_j \in [l_j, u_j])$$

where $m \in Z^+$ specifies the required number of resources and $a_j \in [l_j, u_j], a_j \in A$ specifies the desired range of attribute values ($l_j$ and $u_j$ are lower and upper bounds, respectively). If $l_j = u_j, \forall a_j \in A$, $Q$ is referred to as a point query. Queries that are more complex may also specify Boolean conditions. A hybrid query consists of a set of attributes that may specify a mix of point values, ranges, and Boolean conditions. The set of attributes specified in a query ($Q_A$) may contain only a subset of the attributes for a resource (i.e., $Q_A \subseteq A$). Unspecified attributes are considered as "don't care". $Q$ is referred to as a single-attribute query when $|Q_A| = 1$ and it is referred to as multi-attribute when $|Q_A| > 1$. In rare cases, a query may specify only the required number of resources without specifying any attributes, i.e., $|Q_A| = 0$. Such queries may appear in grid, cloud, and P2P clouds when users are interested in just finding a set of nodes regardless of their capabilities [15]. We term such queries as zero-attribute queries. An example multi-attribute hybrid query may look like follows:

$$Q = \begin{pmatrix} Resources = 6, CPUSpeed \in [2.0\,GHz, MAX], MemoryFree \in [256\,MB, 512\,MB], \\ Architecture = \times 86, OS = "Linux\_2.6.32" \text{ OR } "UNIX\_11.12" \end{pmatrix}$$

In practice, queries are specified using metasyntax notations such as the Extended Backus–Naur Form (EBNF) [16], virtual grid Description Language (vgDL) [17], and XML [14, 16]. When resources are discovered proactively, during the discovery phase, nodes can query the local repository for relevant RSs. Selection of resources needs to take into account both the resource attributes and usage constraints specified in RSs. If a sufficient number of matching resources is not found or a local repository does not exist, nodes can query the centralized database, DHTs, or other peers (whichever supported) in the system.

4. *Match* – Not all the combinations of resources satisfying a resource query may be suitable or capable of working together. It is important to take into account how two resources relate and interact with each other (e.g., bandwidth/latency between different pairs of peers), to ensure that they can satisfy the resource and application constraints. For example, data processing in CASA, GENI, or P2P clouds may not only require processing nodes and storage nodes but also require certain latency/bandwidth bounds among different pairs of such resources. For instance, a node with somewhat limited processing and storage capabilities may be a better match for a certain application than two nodes (one with high processing capabilities and the other with large storage), if they are separated by a low-bandwidth or a heavily loaded interconnect. Similarly, mobile P2P and ad-hoc networks may need to ascertain whether two resources are nearby to avoid a certain service provider, minimize latency, or reduce packet loss. Moreover, social relationships between users may affect the willingness to share their resources. In practice, such constraints on inter resource relationships are expressed as part of the resource query; however, they are evaluated only during the match phase. For example, EBNF notation in [16] and vgDL [17] allow a query to specify groups of resources and then specify intra-group and inter-group constraints. Following is a simplified representation of such a query:

```
Q = {
    Group A
        Resources = 6, CPUSpeed ∈ [2.0 GHz, MAX], DiskFree ∈ [2 GB, MAX]
        Latency ∈ [0, 50 ms]
    Group B
        Resources = 4, CPUFree ∈ [50 %, 90 %], DiskFree ∈ [2 GB, MAX]
        Bandwidth ∈ [2 Mbps, MAX]
    Intergroup
        Latency(A, B) ∈ [0, 200 ms]
        Bandwidth(A, B) ∈ [1 Mbps, 10 Mbps]
}
```

Ability to specify constraints as part of a query enables specific implementations to integrate select and match phases together, and to terminate a range query as soon as the desired set of resources is found. Depending on the types of resource constraints that need to be satisfied by collaborative applications, the P2P system may rely on various inference engines to identify inter-node latencies and bandwidths, network topologies, administrative policies, social relationships, etc.

5. *Bind* – Once a subset of resources that match the application requirements are identified, it is necessary to ensure that the selected resources are available for use. Due to churn or failures, the resources found may not be available by the time the application is ready to utilize them. The same resource may also be under consideration by other applications. Hence, a binding has to be established between the resources and the application trying to use them. Binding is particularly important in guaranteed service environments like CASA and GENI to achieve desired QoS and real-time requirements. Depending on the application environment, a binding may be extended (if a resource(s) is not already committed to another application) without going through the first four phases to reacquire resources. This is useful when an application did not complete its task within the predicted time window or an external event caused the application to continue further (e.g., due to unpredictability of weather events in CASA or increased user demand on a P2P cloud application).

6. *Use* – Utilize the best subset of available resources that satisfy application requirements and constraints to carry out the application tasks for which resources were acquired. Resource usage and interaction patterns are application specific. Users may manually configure simple applications while more complex interaction patterns may be defined as workflows (i.e., logical sequence of actions and interactions among resources). In such applications nodes may collaborate by sharing their resources (e.g., sharing a processing node or a file). Moreover, applications may be also deployed as a set of collaborating agents autonomously moving among selected resources (e.g., in-network data fusion in CASA and bargaining agents on a mobile social network). Resources may be granted for exclusive access or for simultaneous sharing across multiple applications (e.g., prototype CASA radars are shared among concurrent GENI users through virtual machines [18]).

7. *Release* – Release resources when application demand decreases, the task is completed, or binding expires, whichever occurs first. As with usage patterns, the resource release patterns are also application dependent. An application may either release a resource(s) completely (e.g., by removing the application from the resource or terminating the connection) or may reduce its resource usage such that the resource can collaborate with another application (e.g., by reducing CPU load and memory consumption without leaving a processing node). If the binding expires and the application is not able to extend it any further, either the resource(s) or a policy enforcement agent may pull the application out of the resource to ensure that the resource is available to other applications.

Applications may cycle through these phases as and when they need additional resources to fulfill increased/varying application demands, to take advantage of new resources, or to overcome limitations caused, e.g., by resources that fail or leave the P2P network. It is possible for different applications and multiple instances of an application to be in different phases at any given time. Thus, the P2P resource allocation schemes have to continually adapt and evolve based on changing resource availability and dynamic application demands in a robust and scalable manner. The term *resource discovery* typically refers to the first three phases [16, 19-25] whereas the term *resource aggregation* refers to the overall process of advertising, discovering, selecting, matching, and binding resources.

Implementation of these phases is nontrivial, as heterogeneous, multi-attribute, dynamic, and distributed resources and their diverse resource relationships make collaboration quite complex. Nevertheless, a good solution should efficiently advertise all the resources together with their current state, discover potentially useful resources, select resources that satisfy application requirements, match according to application and resource constraints, and bind resources and applications to ensure required QoS guarantees. Moreover, the overall solution needs to be adaptive, fault tolerant, robust, and should satisfy multiple aspects of scalability such as the query resolution latency, number of messages, size of resource index, size of routing tables, and number of attributes. Incentives, trust, privacy, and security have to be an integral part of the phases as they have a direct impact on users' resource contribution and resource collaboration, as well as the long-term stability of a collaborative P2P system. Implementation of these phases depends on the P2P network organization, and below we discuss alternative design choices using a representative subset of P2P solutions. Our objective is to cover the spectrum of alternative design choices rather than provide an in-depth discussion of each solution. These design choices are discussed in the context of phases of P2P resource collaboration; therefore, our discussion is complementary to prior surveys on multi-attribute resource lookup [26-27].

## 3    State of the Art

P2P architectures can be broadly categorized as structured and unstructured overlays [28], the features of which are compared in Table 1. Unstructured overlays are attractive as they are relatively simple to construct and maintain. Moreover, they distribute resource information across many nodes in the system while providing resilience and load balancing. Structured P2P systems are known for high scalability and some guarantees on performance, and hence are utilized in large-scale and relatively robust environments. These systems use a Distributed Hash Table (DHT) to index resources. Each DHT node and a resource have a unique identifier called a *key*. Each resource's contact information is indexed, i.e., stored, at a node having a close by *key* in the key space. A deterministic overlay network is used to route messages that advertise RSs and query for resources. Chord, Kademlia, CAN, and Pastry are some well-known solutions used to build such structured overlays [28]. A representative subset of centralized, unstructured, and structured overlay-based solutions that support different phases of resource collaboration is discussed next.

**Table 1.** Comparison of structured vs. unstructured P2P systems.

|  | **Unstructured P2P** | **Structured P2P** |
|---|---|---|
| Overlay construction | High flexibility | Low flexibility |
| Resources | Indexed locally (typically) | Indexed remotely in a distributed hash table |
| Query messages | Broadcast or random walk | Unicast |
| Content location | Best effort | Guaranteed |
| Performance | Unpredictable | Predictable bounds |
| Object types | Mutable, with many complex attributes | Immutable, with few simple attributes |
| Overhead of overlay maintenance | Relatively low | Moderate |
| Peer churn & failure | Supports high failure rates | Supports moderate failure rates |
| Applicable environments | Small-scale or highly dynamic environments with (im)mutable objects, e.g., mobile P2P | Large-scale & relatively stable environments with immutable objects, e.g., desktop file sharing |
| Examples | Gnutella, LimeWire, Kazaa | Chord, CAN, Pastry, Kademlia |

### 3.1 Centralized Solutions

Though collaborative P2P applications are distributed, to ease the resource aggregation, RSs can be advertised and indexed at a well-known central location within or outside the P2P system. Such indexes are utilized in PlanetLab [29], GENI, grids, and cloud computing. A centralized index can resolve complex multi-attribute range queries by selecting, matching, and binding the best set of resources, as it is aware of all the resources in the system and their constraints. Kee et al. [17] proposed an efficient algorithm that integrates selecting, matching, and binding resources. Furthermore, latency and cost of resource advertising and querying are minimum as nodes can directly communicate with the central node, i.e., cost per message is $O(1)$. However, this approach is not scalable and could lead to a single point of failure. Such failures are not desirable in mission critical systems such as CASA. Hierarchical indexes are proposed to overcome some of these limitations [9] where separate indexing nodes are assigned to different geographic regions, sites, or administrative domains. Then a higher-level node(s) is assigned to keep track of aggregate resources from multiple nodes. This approach could lead to conflicts while querying and binding resources, and partial failures in sub-regions of the system are an issue. Moreover, aggregation along the hierarchy reduces the resolution at which RSs are advertised. Such hierarchies are not desirable while indexing highly dynamic attributes such as free CPU and bandwidth.

## 3.2 Unstructured-Overlay-Based Solutions

Unstructured P2P systems are based on random overlays where resources are advertised, discovered, and/or selected by sending messages through flooding or random walks. Flooding can be used to advertise RSs and/or to select resources on the fly using multi-attribute queries [3]. Either way, all the nodes/queries can get to know about all the resources in the system, which enables the most suitable set of resources to collaborate. However, as flooding is extremely costly, this approach is suitable only for small-scale and/or mobile P2P applications.

Gossiping is another form of disseminating resource information in which RSs propagate in the network due to pairwise exchanges of RSs between randomly selected node pairs [30-31]. Multiple concurrent agents are used to speed up the gossiping. A node selects resources by querying the locally stored RSs that are gathered from gossiping. A node can perform simple resource matching within itself as it gets to know about multiple resources, e.g., "*are resources x and y from my friends?*". Though this approach is simple to implement, there is no guarantee that a node will get to know about the resources that it needs even when they exist. Moreover, the state of knowledge about resources may be stale (e.g., state of dynamic attributes such as free CPU, memory, and bandwidth), as gossip propagation is unpredictable and slow.

Majority of the unstructured P2P solutions are based on random walks where RS advertisements, resource discovery probes, and/or multi-attribute resource queries are forwarded from node to node via neighboring nodes selected randomly. Random walk is a specific form of gossiping where an agent moves from one node to another (randomly selected) carrying a selected set of RSs or multi-attribute queries. In [19], a node sends out multiple agents to advertise its RSs to other nodes along their random walks and to collect RSs of nodes being visited. Multiple agents are sent to different regions of the network to sample larger portions of the network and enhance the robustness. Simple resource matching is also possible as agents collect RSs from multiple nodes. Though agents advertise and discover resources, they do not provide guaranteed resource selection due to random sampling that occurs. Moreover, the state of the selected resources may be stale by the time they are utilized. Same applies to agents that are sent to discover potentially useful resources without explicit resource advertisements. Alternatively, inspired by the second-generation Gnutella P2P system, many solutions issue queries as and when they need to select resources. Each query agent walks from one node to another random node looking for resources specified in the query. When the relevant resources are found, the agent goes back to the query initiator directly, or via the reverse path which helps to anonymize communication between the resource provider and consumer. Agents have a limited lifetime, defined in terms of maximum number of hops ($hops_{max}$), to prevent the accumulation of unresolved query agents within the P2P system. Thus, the cost of resource advertising or querying is $O(hops_{max})$ per agent. To increase the probability of resource selection a node sends either several agents or one agent with a higher lifetime, i.e., large $hops_{max}$. However, it is nontrivial to determine the optimum number of agents or $hops_{max}$ to guarantee resource selection in an arbitrary overlay network. Therefore, such solutions can only provide a best-effort resource selection service. In [32], we observed that random-walk-based solutions can perform better (lower $hops_{max}$ with moderate query hit rate) under real workloads which tend to define less specific queries with a few attributes and large range of attribute values. However, a very large $hops_{max}$ value is still needed to further improve the query-hit rate. Nevertheless, a query agent can identify the current state of a resource and provide resource binding as it reaches individual nodes to check their resource availability. Resource matching can also be performed at the same time, if a node has multiple resources (e.g., if a node has processing capabilities and storage, it can match a query searching for a processing node and a storage node with a given inter-node bandwidth/latency. Past-query results can be used to bias the random walk towards potential resources while reducing the query overhead and latency. However, such optimizations are only applicable to static or slowly varying dynamic attributes such as CPU speed, operating system version, and free disk space [15]. Another alternative is to build a hybrid system where one set of agents advertises the RSs while another set queries for required RSs. While this speeds up the query resolution, it may not reflect the correct state of dynamic resources and also eliminates the possibility of resource binding as queries are answered by intermediate nodes.

Another alternative is a two-layer overlay (similar to Kazaa [28]) where resource-rich peers, namely *superpeers*, form a separate overlay while acting as proxies for rest of the peers [20, 30-31]. Each superpeer keeps track of RSs of a subset of peers. Superpeers advertise, discover, and/or select resources on behalf of their peers by contacting other superpeers through flooding [20], gossips [30-31], or random walks. This approach enhances multi-attribute query resolution, as superpeers are aware of multiple RSs. Moreover, overhead is reduced as only the superpeers are involved in advertising, discovering, and selecting resources. In [32], we observed that superpeers could resolve real-world queries with relatively high hit rate under moderate overhead. Advertising cost can be further reduced by advertising only the aggregate information about resources indexed at a superpeer [30], e.g., total number of CPU cores or total processing capacity (measured in FLOPS) of all the peers assigned to a superpeer. Though such aggregated attribute values are acceptable for desktop grids and bag-of-task applications [30], they are too abstract for latency sensitive applications such as CASA, mobile social networks, and P2P clouds where performance is determined by the interactions among dynamic and individual resources. Nevertheless, superpeers can provide resource binding on behalf of their peers. Moreover, they have the potential to act as matchmakers when the peers in the same neighborhood or with different resources are assigned to the same superpeer.

Unstructured P2P solutions provide a best-effort service for discovering, selecting, and binding resources. They are applicable in small to medium scale applications and in highly dynamic environments such as ad-hoc and mobile social networks. Issuing queries on the fly is more useful because multi-attribute range queries can be easily resolved as agents reach individual nodes. This further enables resource binding. Depending on the specific implementation, a restricted form of resource matching is also possible. If a node cannot match all the relevant resources by itself, information on all the resources that satisfy the resource selection phase has to be provided to the application that initiated the query. In the worst case, this may require sampling all the resources in the system, as what resources may match cannot be determined a priory. The application then takes the final decision on resource matching and binding. Nevertheless, random topologies in unstructured P2P systems make it hard to keep track of complex inter-resource relationships. Therefore, resource relationships have to be discovered as required. For example, after selecting potential resources, an application may request resources to verify whether they can reach each other and the node initiating the application (e.g., by sending ping messages), measure bandwidth/latency between them, or evaluate their social relationships. However, discovering such relationships on the fly takes time and increases overhead, e.g., sufficient number of packets and time are needed to estimate the bandwidth between a pair of nodes. Superpeers provide a viable alternative where a superpeer can at least keep track of the relationships among resources that are assigned to it.

### 3.3 Structured-Overlay-Based Solutions

Table 2 compares alternative design choices available for structured P2P systems (see [28] for details). Let us briefly discuss Chord, as several resource discovery solutions under discussion depend on it. Chord maps both peers/nodes and resources into a circular address space, namely a *ring,* using consistent hashing. A node is assigned to a random location within the ring and is responsible for indexing keys that map into its contiguous range of addresses. A resource's identifier, e.g., file name, is hashed to produce a *key*, and the (*key*, *value*) pair is then stored at its *successor* (i.e., closest peer in the clockwise direction). *value* can be either the resource itself (e.g., file) or its contact details (e.g., IP address and port number of the node having the resource). Each peer maintains a set of pointers, known as *fingers*, to peers that are at an exponentially increasing distances within the ring. These fingers are used to recursively forward a query to its successor within $O(\log n)$ hops, where $n$ is the number of nodes in the overlay. Chord and other structured P2P solutions are designed to index objects that are characterized by a single attribute. Thus, they are not efficient for simultaneous selection of multi-attribute resources required by collaborative P2P applications. Next, we discuss several solutions that extend/modify structured P2P systems to support multi-attribute resources.

**Table 2.** Summary of structured P2P solutions.

| Scheme | Architecture | Routing mechanism | Lookup overhead*† | Routing table size* | Join/leave cost† | Resilience† |
|---|---|---|---|---|---|---|
| CAN | *d*-torus | Greedy routing through neighbors | $O(dn^{1/d})$ | $2d$ | $2d$ | Moderate |
| Chord | Circular key space | Successor & long distance links | $O(\log n)$ | $O(\log n)$ | $O(\log^2 n)$ | High |
| Cycloid | Cube connected cycles | Links to cyclic & cubical neighbors | $O(d)$ | $O(1)$ | $O(d)$ | Moderate |
| Kademlia | Binary tree, XOR distance metric | Iteratively find nodes close to key | $O(\log n)$ | $O(\log n)$ | $O(\log n)$ | High |
| Mercury | Circular key space | Successor, predecessor, & long distance links | $O(1/k \log^2 n)$ | $k + 2$ | $k + 2$ | Moderate |
| Pastry | Hypercube | Correct one digit in key at time | $O(\log_B n)$ | $O(B \log_B n)$ | $O(\log_B n)$ | Moderate |
| Tapestry | Hypercube | Correct one digit in key at time | $O(\log_B n)$ | $O(\log_B n)$ | $O(\log_B n)$ | Moderate |
| Viceroy | Butterfly network | Predecessor & successor links | $O(\log n)$ | $O(1)$ | $O(\log n)$ | Low |

\* $B$ – base of a key identifier, $d$ – number of dimensions, $k$ – no of long distance links, $n$ – number of nodes in overlay

† Lookup overhead – number of overlay hops required to resolve a point query, Join/leave cost – number of overlay maintenance messages required to stabilize the network after the arrival/departure of a node, Resilience – ability to deliver messages under node failure.

Figure 3 illustrates three design choices for the case of multi-attribute resources based on a ring-like overlay network. Mercury [21] maintains a separate ring for each attribute type (see Fig. 3(a)). Each node advertises its RS(s) to rings that correspond to the attribute set of its resources. Instead of consistent hashing used in Chord, a Locality Preserving Hash (LPH) function is proposed. LPH functions map close by attribute values to neighboring nodes in the ring enabling range query resolution through a series of successors (Fig. 3). This enables a range query $Q$ to be resolved by forwarding it to a series of nodes through their successor (see Fig. 3(a)). Given a range of attribute values $[l, u]$, $Q$ is first routed to the peer responsible for indexing $l$. $Q$ is then forwarded through successor pointers/fingers until the peer responsible for indexing $u$ is reached or the desired number of resources is found, whichever comes first. Thus, the number of hops required to resolve a range query is proportional to the number of nodes $n$ in the ring. Therefore, query resolution cost is bounded by $O(n)$ [15, 32]. Mercury utilizes single-attribute-dominated queries where $Q$ is issued only to the ring corresponding to the *most selective attribute*. For example, as seen in Fig. 3(a), the query travels 3-hops in the CPUSpeed and bandwidth rings while it travels only 2-hops in the memory ring. Therefore, $Q$ is issued only to the memory ring, which is the most selective (i.e., minimum query resolution cost) attribute for the given query. To support such queries, each node keeps track of all attributes of RSs that it indexes and uses a random-sampling algorithm to estimate the query selectivity within each ring. Therefore, the index size of a node is bounded by $O(RA)$, where $R$ is the number of resources in the system and $A$ is the number of attributes used to describe a resource. Even if one of the attributes of a resource changes, it has to be advertised to all the relevant rings. Therefore, advertising cost can be significant when the collaborative P2P system has a large number of dynamic attributes that change rapidly [15, 32]. Mercury can add new rings to support additional attributes with minimum modifications. However, a large number of routing entries associated with multiple rings makes it inappropriate for collaborative P2P systems such as CASA and GENI that need to support a diverse set of resources characterized by tens to hundreds of attributes. In [15], it was observed that the multi-attribute queries in real-life systems are skewed, i.e., a small set of attributes and attribute value ranges were highly popular. Such a skewed distribution of queries distributes the load among rings unequally, with the segments of the rings corresponding to these popular ranges of attribute values getting highly utilized while the majority of the ranges/rings are underutilized.

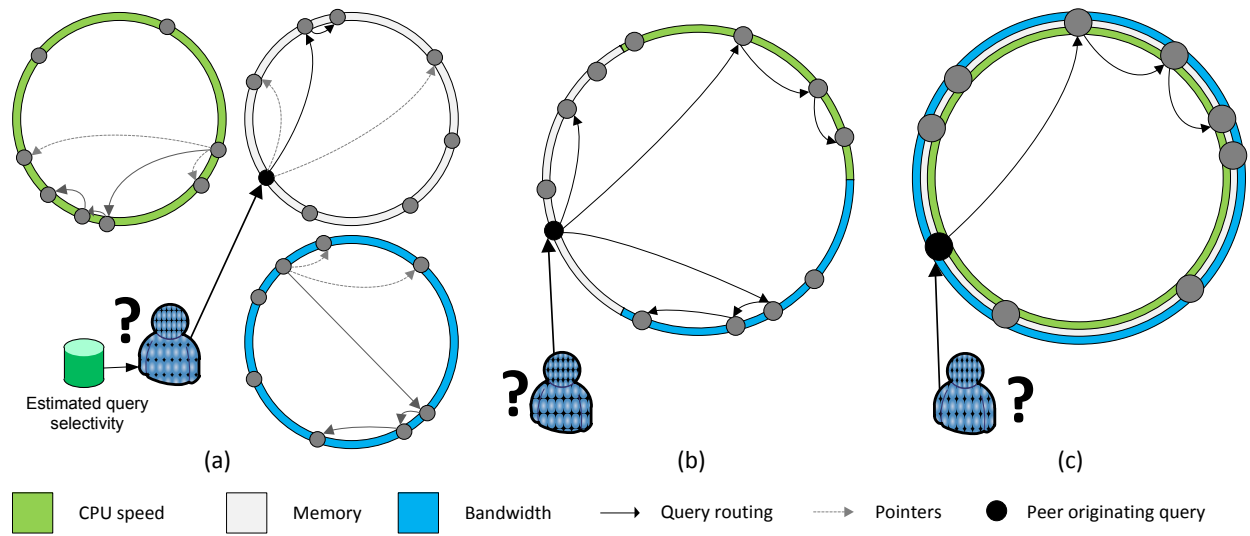

**Figure 3.** Ring-based structured overlay designs: (a) Separate overlays for each attribute type and single-attribute-dominated query resolution; (b) Overlay partitioned to accommodate multiple attribute types and query resolution using multiple sub-queries; (c) Overlay with overlapped attribute space and single-attribute-dominated query resolution.

Figure 3(b) illustrates an alternative design based on a partitioned ring. LORM (Low-Overhead, Range-query, and Multi-attribute) [22] assigns a different segment of the ring to each attribute type. Attribute values are represented as a bit string where prefix bits represent the segment in the ring (i.e., attribute type) and suffix bits represent the position within a segment. Suffix bits are generated by applying a LPH function to the attribute value. An application searching for resources issues a separate sub-query for each segment according to the required set of attributes specified in $Q$.

Query results are then combined at the application using a join operation like that in databases. MADPastry (Mobile AD-hoc Pastry) [33] proposes a similar scheme for large-scale mobile ad-hoc networks. It segments the ring to preserve the physical locality of nodes determined using a set of landmarks. Nodes that are close to the same landmark have identical prefix bits. Sub-queries are first sent to the local segment. If the required resources are not found locally, sub-queries are then sent to other segments. Another alternative is to build a set of hypercubes for each geographic region and then connect them to form a backbone [34]. These solutions maintain a much lower number of routing entries compared to Mercury. MADPastry and hypercube also provide resource matching based on latency and hop count because of their ability to locate local resources. Although one of the attributes satisfies a sub query, rest of the attributes of the resource may not satisfy the application requirements. Therefore, most of the resources returned by a sub-query are not useful. Consequently, multiple sub-queries and their tendency to return a large number of unsuitable resources increase the lookup overhead in all three solutions. Moreover, LORM also suffers from load imbalance where skewed distribution of attributes in real workloads [32] force several segments to be highly utilized while remaining segments are rarely utilized. MADPastry and hypercube cannot be used in highly dynamic mobile ad-hoc networks, as the structured overlay topology can be easily disconnected/partitioned due to high mobility of peers.

MAAN (Multi-Attribute Addressable Network) [23] proposes a single-attribute-dominated query mechanism for a ring (Fig. 3(c)). It maps attribute values of a resource to the same address space using a separate uniform LPH function for each attribute type. In addition to preserving the locality, uniform LPH functions also uniformly distribute hash values across the ring thus providing static load balancing. Though use of an overlapped rings reduces the routing table size, the number of overlay hops required to resolve a range queries is still $O(n)$ [15, 32]. Moreover, uniform LPH functions fail to balance the load when resource queries are skewed or when there are many identical resources, which is known to be the case in production systems such as grids, desktop grids, clouds, and CASA [15, 32].

Figure 4 shows another design where RSs are mapped to points in a $d$-dimensional torus where each dimension represents one attribute type. MURK (Multi-dimensional Rectangulation with Kd-trees) [24] is one such solution that dynamically partitions the torus and assigns each partition to a separate node. Each node indexes all the RSs that map into its own partition. MURK keeps track of these partitions by organizing them in the form of a $k$-dimensional tree. Dynamic load balancing is achieved by splitting/aggregating the partitions based on the query load. A multi-attribute range query $Q$ encloses a contiguous region (i.e., hyperrectangle) on the torus, e.g., $Q_1$ in Fig. 4(a). Real-world queries specify only a few attributes [15]; therefore, volume of the hyperrectangle can be extremely large due to many "don't care" attributes. Queries are resolved using greedy forwarding, which forwards $Q$ to each partition that overlaps with the query region. As $Q$ propagates, each partition reports the selected resources to the application that initiated the query, as it is not straightforward to aggregate resources from multiple partitions. As each partition is not aware of what resources were selected by other partitions, $Q$ has to visit all the partitions that overlap with the query region even though enough resources may have been already found. Alternatively, MURK uses space-filling curves [24] to map the $d$-torus to a Chord ring while reducing its dimensionality. Chord ring enables $Q$ to aggregate potential resources as it propagates and terminates $Q$ as soon as the desired number of resources is found. It introduces additional overhead as nearby resources on the $d$-torus are no longer mapped to a contiguous region on the ring. Hence, query resolution cost is bounded by $O(n)$. Moreover, space-filling curves loose locality when extended to a large number of dimensions.

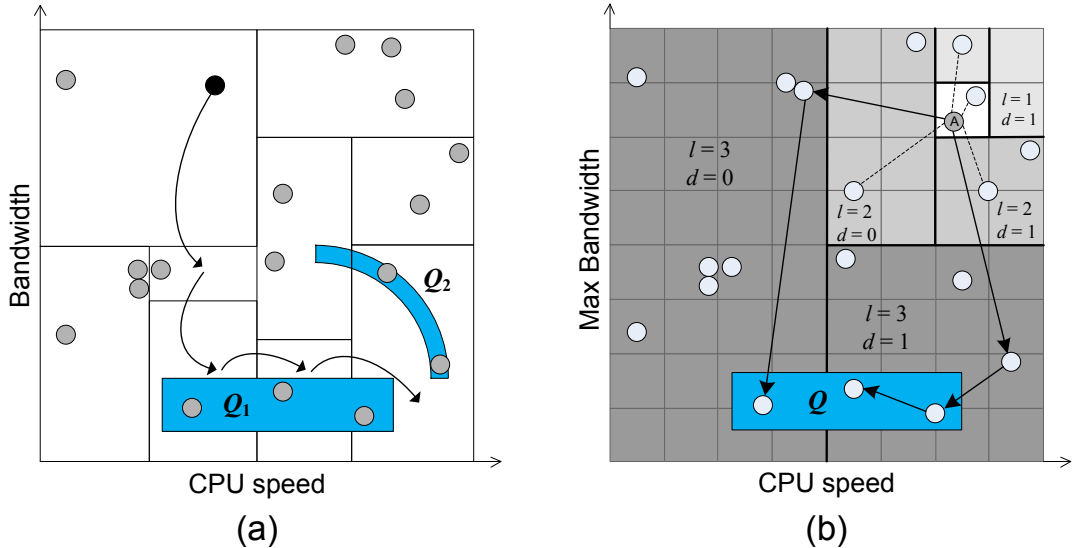

**Figure 4.** 2-dimensional torus: (a) Multi-attribute space partitioned among peers, and a query is forwarded to all the partitions that overlap with the query region; (b) Multi-attribute space partitioned with respect to peer *A*, query is forwarded to all the peers that overlap with the query region, *l* – level, *d* – dimension [25].

Mercury, LORM, MAAN, and MURK provide only resource advertisement and selection (resource discovery is not required as resources are explicitly advertised). SWORD is a partitioned-ring-based architecture (see Fig. 3(b)) that also provides resource matching [16]. A query in SWORD can specify a range of attribute values for each required resource, how to group the resources, intra-group and inter-group resource relationships/constraints, and penalties for not satisfying them. As it is impractical to keep track of all the inter-resource relationships, SWORD clusters resources/nodes into equivalence classes based on their Autonomous System (AS). A designated node from each AS keeps track of latency/bandwidth relationships between the ASs. Resources are either selected using a single-attribute-dominated query (as in MAAN) or multiple sub-queries (as in LORM). The selected resources are then used to form a set of *candidate groups* according to the application requirements. Subsequently, resources in each candidate group are matched based on inter-AS relationships. The groups are then ranked according to the extent they satisfy application constraints and sent to the application allowing it to take the final decision on which group(s) to use. AS-level measurements are too coarse grained and may not work well for bandwidth and other complex resource relationships, which are required in CASA, GENI, and P2P clouds.

Solutions discussed so far are applicable in semi-dynamic environments such as Grid Computing where resources do not change rapidly. Costa et al. argue that DHTs are inefficient and incapable of accurately maintaining the state in highly dynamic environments such as P2P clouds [25], and propose to construct a *d*-torus using only the static attributes that can be represented correctly (see Fig. 4(b)). In contrast to other structured P2P solutions, resources are not explicitly advertised and indexed remotely. Instead, the overlay network is formed by maintaining pointers to nodes with different attribute values as seen in Fig. 4(b), which are identified using gossiping. The torus is recursively partitioned into smaller and smaller hypercubes called *cells*. A query is first routed to one of the lowest-level cells, using the pointers maintained to other nodes that overlap with the query region. Other overlapping cells are then recursively traversed using depth-first search until all the required or available resources are found. This approach can advertise, select, and bind resources as queries are forwarded to individual nodes where the availability of resources and dynamic attributes defined in a query can be evaluated. However, depth-first search significantly increases the lookup overhead and latency, particularly under less specific queries in real workloads [15, 32]. In [32], we also observed that majority of the queries in real-life systems specify only the dynamic attributes. This solution cannot be used in such cases, as at least one static attribute is needed to initiate the query resolution or entire $d$-torus has to be searched.

Table 3 summarizes the different structured P2P solutions, and all the major solutions are compared in Table 4 with respect to advertising, discovering, selecting, matching, and binding resources. There is no universal solution and each has its own distinct advantages and limitations. In [32], we evaluate the performance of seven of these basic overlay designs by simulating traces of multi-attribute resources and queries derived from PlanetLab [16, 29]. Results show that structured and unstructured P2P designs have a much higher advertising and querying cost than anticipated in their original publications, and suffer from load balancing issues where few nodes indexed most of the RSs and answered most of the queries. Compared to other solutions, the cost of resolving single-attribute-dominated queries is relatively lower [32]. However, those solutions also have higher advertising cost, unbalanced query load, large resource indexes, and tend to inaccurately representation state of dynamic resources due to the large number of replicas [15, 32]. Sub-query-based solutions can overcome most of these issues but they have a significantly higher query cost. Mercury, MAAN, LORM, and SWORD architectures are extensible because they can add new attributes without significant modifications to the overlay. Except for SWORD, MADPastry, and hypercubes, the other solutions do not support any form of resource matching. Resource-aware overlay [25] is the only structured P2P solution that supports resource binding. However, it also has high query cost and cannot resolve queries with only the dynamic attributes. We believe these findings are also applicable to many recent solutions that have been proposed based on these alternative design choices. Moreover, many recent solutions also do not address the problem of resource matching or binding.

**Table 3.** Comparison of structured P2P solutions.

| Scheme | Architecture | Routing mechanism | Lookup overhead (point query)* | Lookup overhead (range query)* | Routing table size* | Load balancing |
|---|---|---|---|---|---|---|
| Mercury [21] | Multiple rings | Successor, predecessor & long distance links | $O(1/_k \log^2 n)$ | $O(n)$ | $k + 2$ per ring | Dynamic |
| LORM [22] | Partitioned ring | Cycloid | $O(d)$ | $O(n)$ | $O(1)$ | Static |
| MADPastry [33] | Partitioned ring (locality based) | Pastry | $O(\log n)$ | $O(n)$ | $O(\log l)$ | Static |
| Hypercube backbone [34] | Hypercube-based backbone | Hypercube | Local – $O(d)$ Remote – $O(D)$ | $O(n)$ | $\Theta(d)$ | Dynamic |
| MAAN [23] | Single ring | Chord | $O(\log n)$ | $O(n)$ | $O(\log n)$ | Static |
| MURK [24] | $d$- torus | CAN with long distance links | $O(\log^2 n)$ | $O(n)$ | $2d + k$ | Dynamic |
| SWORD [16] | Partitioned ring, resource matching | Chord | $O(\log n)$ | $O(n)$ | $O(\log n)$ | Static |
| Resource-aware overlay [25] | $d$- torus partitioned into cells | Links to peers in other cells | $O(n)$ | $O(n)$ | $O(d)$ | Static |

\* $n$ – number of peers in overlay, $k$ – number of long distance links, $d$ – number of dimensions, $D$ – network diameter, $l$ – number of landmarks

**Table 4.** Summary of all the solutions with respect to resource advertise, discover, select, match, and bind phases.

| Scheme | Architecture | Advertise | Discover* | Select | Match* | Bind* |
|---|---|---|---|---|---|---|
| Flooding [3] | Flood advertisements or queries | Yes | N/A | Guaranteed | When RSs are flooded | When queries are flooded |
| Gossiping [30-31] | Agents share resource specifications they know | Yes | Yes | Moderate probability of success | Simple matching | N/S |
| Random walk [19] | Agents carry RSs & queries | Yes | Yes | Moderate probability of success | Simple matching | When query agents are used |
| Superpeer [20, 30-31] | 2-layer unstructured overlay | Yes | Yes | Relatively high probability of success | Simple matching | Yes |
| Mercury [21] | Multiple rings | Yes | N/A | Guaranteed | N/S | N/S |
| LORM [22] | Partitioned ring | Yes | N/A | Guaranteed | N/S | N/S |
| MADPastry [33] | Partitioned ring (based on locality) | To local & neighbor partitions | N/A | Guaranteed | Latency & hop count | N/S |
| Hypercube backbone [34] | Hypercube based backbone | To local & neighbor hypercubes | N/A | Guaranteed | Latency & hop count | N/S |
| MAAN [23] | Single ring | Yes | N/A | Guaranteed | N/S | N/S |
| MURK [24] | *d-* torus | Yes | N/A | Guaranteed | N/S | N/S |
| SWORD [16] | Partitioned ring, resource matching | Yes | N/A | Guaranteed | Latency & bandwidth | N/S |
| Resource-aware overlay [25] | *d-* torus partitioned into cells | Static attributes only | N/A | Guaranteed | N/S | Yes |

\* N/A – Not applicable, N/S – Not supported

## 4 Incentives, Trust, Privacy, and Security

P2P systems are designed to be open, dynamic, collaborative, and perhaps even anonymous, and hence are vulnerable to abuse by selfish or malicious users. Success of a collaborative P2P system depends on the extent to which users contribute their resources for the common cause. Hence, proper incentives need to be in place to motivate users to contribute their resources. However, users will continue to contribute their resources only if they feel that other users are trustworthy and are also contributing to the common cause without trying to game the system for their own advantage. Furthermore, users also need to feel safe that their accumulated incentives, gained trust, privacy, and security will not be compromised while they continue to contribute resources. Therefore, incentives, trust, privacy, and security are related and need to be consistently enforced throughout all the key phases of resource collaboration. Next, we briefly discuss a representative subset of solutions that have been proposed to address these concerns.

Incentives need to be embedded into a collaborative P2P system as they encourage resource contribution and collaboration while enabling the applications to accomplish their tasks with sufficient performance and quality. Incentives may be based on monetary or nonmonetary benefits. For example, Napster and BitTorrent utilize various Web 2.0 technologies to build online communities of users around common interests [35]. Members of a community may altruistically contribute their resources for the satisfaction of being in a community and for the recognition gained by becoming an active contributor. Similar incentives can work for collaborative P2P applications such as CASA, crowd-sourced weather monitoring, and mobile social networks deployed during a rally or after a major disaster. However, Zhao et al. [36] showed that there is a tradeoff between altruism and system robustness, and analytically justified that a P2P system should limit the degree of altruism to encourage cooperation. To this end, several monetary-benefit-based incentive schemes (e.g., micropayments and scores) are proposed. Golle et al. presented several micropayment schemes for P2P file sharing and demonstrated their efficacy using game theory [35]. Furthermore, there is evidence that private BitTorrent communities already use various scoring schemes to keep track of user contributions in terms of bytes or number of files uploaded [10-11]. Members of such communities tend to take cost conscious decisions while deciding what files to upload and download [10]. This approach can be extended to collaborative P2P systems that rely on a centralized index, which also keeps track of micropayments/scores during resource bind or release phases. However, centralized indexes are not scalable, prone to attacks, and cannot work with mobile social and ad-hoc networks. Alternatively, a distributed payment scheme based on a DHT is presented in [37]. [37] represents a node $n$'s resource contribution/consumption as an increasing/decreasing scalar value $v_n$. $v_n$ is replicated on a set of DHT nodes called *bankers* (e.g., $k$-nearest neighbors of $n$ in the DHT). Multiple replicas are maintained to increase resilience and minimize the possibility of all bankers being attacked simultaneously. This approach can be extended to large-scale P2P clouds, CASA, and GENI, which may utilize a DHT-based resource discovery system.

Collaborative applications need to be able to find trustable resources that can collaborate with each other to accomplish the desired end result. Trust relationships among resources may be developed based on credentials, reputation, and social connections of resource owners. Credential-based solutions establish a binary trust (i.e., trust completely or not at all) relationship between a pair of resources based on some credential such as a digital certificate or a shared secret. Such binary trust relationships are primarily used to enforce security and access control to resources. They are appropriate in applications such as CASA and GENI, which consists of different classes of end users, and are administered by specific organizations. Another measure of trust is the mutual reputation that resources place on each other based on their past interactions. Reputation can be defined in terms of the accuracy of data generated by a sensor, reliability of computed results, availability of resources, ability to meet deadlines, etc. When two resources interact with each other, they establish a local trust/reputation value (both resource provider and consumer may rate each other). Let $r_{Local}(i,j)$ be the reputation that resource $j$ places on resource $i$. These $r_{Local}(i,j)$ values are then aggregated to build a global trust value $r_{Global}(i)$ of a resource. $r_{Global}(i)$ values are useful in predicting how two resources that never interacted in the past will behave in the future. While the calculation of $r_{Local}(i,j)$ is application and resource specific, several solutions have been proposed to calculate $r_{Global}(i)$. In [38], $r_{Global}(i)$ is calculated by aggregating $r_{Local}(i,j)$ values provided by a predefined set of highly trusted resources. This approach is not suitable for P2P systems with high churn as failure of highly trusted resources is catastrophic. Alternatively, [39] proposes a mechanism to dynamically identify highly trusted resources within the overlay network by arranging them on a DHT using a locality preserving hash function. Both [38-39] do not provide good incentives for resource collaboration, as resources are motivated to provide services only to the set of highly trusted resources to gain better reputation. Moreover, a set of highly trusted resources may potentially become a single point of failure as they may become overloaded or an easy target for attackers. This problem can be overcome by weighting each $r_{Local}(i,j)$ value by $r_{Global}(j)$ of the nodes providing feedback (i.e., $r_{Global}(i) = \sum_{\forall j} r_{Local}(i,j) \times r_{Global}(j)$)

[40]. An alternative design in [41] assumes that all the resources behave selfishly and represent reputation in terms of a

virtual currency where a resource's reputation is reflected by its wealth. Moreover, it forms groups of resources that trust each other and group members work together to protect their group trust against malicious resources. Each group keeps track of trust values of all the other groups in the system. [40-41] can be used with both structured and unstructured overlays. However, it is not straightforward to maintain a consistent view of trust values among nodes and groups without a costly synchronization or update scheme. Moreover, it is not known whether [40] converges to a stable set of trust values [39]. Trust values can be updated during the resource binding and release phases and can be evaluated when selecting and matching resources. Bootstrapping reputation-based trust systems is also an issue as a sufficient number of local trust values need to be accumulated before a reliable global trust value can be calculated. This is particularly a problem in short-lived mobile social and ad-hoc network applications. In such cases, social networks among resource owners can be taped in as an initial estimation of trust. Collaborative applications that have access to the Internet can tap into online social networks through standard APIs that are being exposed by social network providers. Such solutions can evaluate thrust based on whether resources belong to friends or friends of friends, and so on. When access to online social networks is not possible, resources may start with a random set of known resources as witnesses, e.g., physically close by mobile phones [42]. When two resources that do know trust each other want to collaborate, they look for at least one common witness to establish the trust between them. It has been shown that this approach works well in fast-mixing, scale-free networks [42]. However, not all collaborative P2P systems are scale free and a large number of initial witnesses are needed to guarantee that any two resources can find a common witness with a high probability.

Privacy of resource provider is also important, and monitoring by third parties of their resource contribution and usage patterns may discourage the participation. Moreover, secure communication may be also required among resources collaborating on a given application. While public key infrastructure can facilitate secure channel establishment within applications such as CASA, GENI, and P2P clouds, shared keys and key pre-distribution may be desirable in mobile social and ad-hoc networks. Privacy preservation in P2P systems is mainly accomplished through anonymity where messages are routed through multiple hops in the overlay network preventing resource provider, consumer, and intermediate nodes from knowing who is talking to whom. However, anonymity is a problem when it comes to enforcing incentives and assessing trust among resources as micropayments and trust values are associated with resource identifiers. Moreover, if security enforcement is required to make resources accountable for their actions, it may have to be based on a globally unique identifier. Choi et al [43] propose to overcome this problem by replacing user identity with a temporary network address and a set of pseudonyms approved by a Certifying Authority (CA). CA mediates all initial communication requests among resources and provides authenticity and accountability for resources. While this solution prevents intermediate nodes from spying on each other, certifying authority can now spy on all the resources. The same issue persists with P2P resource aggregation solutions based on a centralized index. Thus, certifying authority not only becomes a single point of failure but also may demotivate resource collaboration, particularly in P2P cloud and mobile social networks. Furthermore, how anonymity and privacy are implemented on a particular collaborative P2P system has an impact on all the phases of resource collaboration as each phase keeps track of resources by their identifiers and RSs.

Open access provided by P2P systems make them easy targets for attacks compared to other distributed systems. Overlay routing tables and resource indexes can be easily compromised to divert applications to malicious resources to initiate man in the middle or Denial of Service (DoS) attacks. Peers may also pick specific keys to place themselves on desired places in a DHT to provide bogus answers to queries. It has been proposed to use a CA to assign overlay keys and provide digital certificates to resources, which can be used to sign RSs and overlay route advertisements. However, verification of certificates and key revocation is not straightforward unless direct access to CA(s) is available, which may not be possible in contexts such as mobile social and ad-hoc networks. Other forms of DoS attacks can be initiated by overloading the resource indexes with many RS advertisements and less specific queries requesting a large number of resources and the entire domain of attribute values. Solutions to such issues may involve rate limiting RS advertisements and queries and forcing users to specify queries that are more specific. Furthermore, resource contributors should be protected from being victims of malicious resource consumers. Otherwise, the entire collaborative P2P system will easily become a large botnet. Alternatively, collaborative applications should be protected from malicious or compromised resources. Such attacks can be minimized using techniques such as sandboxing, virtual machines, and service oriented architecture that exposes services through a well-defined API [44]. For example, prototype CASA radars are exposed to concurrent GENI users through virtual machines [18]. Table 5 groups the different solutions based on the tradeoffs made and their strengths and weaknesses.

**Table 5.** Solutions for P2P resource collaboration grouped according to tradeoffs made and their strengths and weaknesses.

| Schemes | Applicable For | Strong Aspects* | Weak Aspects* |
|---|---|---|---|
| Centralized | Small to medium scale, static networks<br>Guaranteed resource discovery<br>Centrally administered<br>Applications<br>• Single site grid & cloud computing<br>• Desktop grids | Complex & mutable resources<br>All phases of resource collaboration<br>$O(1)$ advertise & query cost<br>Extendable to support new attributes<br>Track & enforce incentives & trust<br>Act as a CA to enhance security<br>No routing state | Single point of failure<br>$O(R)$ index<br>$O(Q)$ query load<br>Lower privacy & trust as centralized node is the only source of trust |
| Hierarchical | Medium to large scale, geographically distributed, static networks<br>Guaranteed resource discovery<br>Multiple administrative domains<br>Applications<br>• Multi-site grid & cloud computing<br>• Desktop grids | Complex & mutable resources<br>All phases of resource collaboration<br>$O(1)$ advertise & query cost<br>Extendable to support new attributes<br>Track & enforce incentives & trust<br>Act as a chain of CAs to enhance security<br>$O(1)$ routing state | $O(R)$ resource index at higher level nodes or resolution degradation if aggregated<br>Local failures & partitioned hierarchy<br>Resource select, match & bind require collaboration across multiple nodes in hierarchy<br>Lower privacy<br>Chain of trust is broken if higher-level nodes are compromised |
| Unstructured + Flooding | Small scale, static/dynamic networks<br>Guaranteed resource discovery<br>Highly robust<br>Applications<br>• Mobile social networks<br>• Mobile ad-hoc networks | Complex & mutable resources<br>All phases of resource collaboration<br>Extendable to support new attributes<br>Privacy through anonymity<br>Highly robust against node failures<br>$O(1)$ resource index<br>Simple to build & maintain network<br>Preserve locality in mobile networks | Very high advertise & query cost – message implosion<br>Match resources only if RSs are flooded<br>Nontrivial to enforce incentives, security, & track trust<br>$O(n)$ routing state |
| Unstructured + Gossiping or Random walk | Small to medium scale, static/dynamic networks<br>Best effort resource discovery<br>Applications<br>• Mobile social networks<br>• Ad-hoc networks<br>• P2P clouds<br>• Desktop grids | Complex & mutable resources<br>All phases of resource collaboration<br>Extendable to support new attributes<br>Privacy through anonymity<br>Highly robust against node failures<br>$O(1)$ resource index<br>Simple to build & maintain network<br>Better load distribution – no explicit | Very high advertise & query cost<br>Not guaranteed to find resources<br>Limited ability to match resources<br>Bind resources only if query agents are used<br>Nontrivial to enforce incentives, security, & track trust<br>Stale resource specifications |

| | File sharing | load balancing necessary<br>Preserve locality in mobile networks | $O(n)$ routing state |
|---|---|---|---|
| Superpeer + Flooding, Gossiping, or Random walk | Medium scale, static or semi-dynamic networks<br>Best-effort resource discovery<br>Applications<br>• P2P clouds<br>• Desktop grids<br>• File sharing | Complex & mutable resources<br>All phases of resource collaboration<br>Extendable to support new attributes<br>Relatively simple to build & maintain<br>Relatively low advertise & query cost<br>$O(1)$ advertising cost<br>Privacy of peers through anonymity<br>Ability to enforce incentives, trust, & security – superpeers can act as CAs<br>Robust against node failures | Not straightforward to pick a superpeer<br>Not guaranteed to find resources<br>Superpeers can monitor peers & are the only sources of trust<br>$O(n)$ resource index<br>$O(n)$ routing state |
| MADPastry & Hypercube backbone | Medium to large scale, static or slowly moving networks<br>Guaranteed resource discovery<br>Applications<br>• Mobile social networks<br>• Ad-hoc networks | Mutable resources<br>Resource selection & match<br>Physical locality somewhat preserved<br>$O(\log n)$ point query cost<br>Match resources based on locality<br>Explicit static & dynamic load balancing – Hypercube only<br>Privacy through anonymity<br>Ability to enforce incentives & trust | Single attribute resources only<br>$O(n)$ range query cost<br>High cost of advertising mutable resources<br>No resource bind<br>Match limited to locality<br>Lacks security unless all nodes have access to a CA |
| Resource-aware overlay | Large scale, relatively static networks<br>Immutable attributes<br>Guaranteed resource discovery<br>Applications<br>• Desktop grids | Complex & immutable resources<br>Guaranteed performance<br>No RS advertise cost<br>Select & bind resources<br>$O(A)$ routing state<br>Static load balancing<br>Privacy through anonymity<br>Track & enforce incentives & trust | No mutable resources<br>$O(n)$ range query cost<br>No resource match<br>No dynamic load balancing<br>Moderate resilience to churn<br>Lacks security unless all nodes have access to a CA |
| Ring-like overlays: Mercury, LORM, MAAN, MURK, SWORD | Large scale, relatively static networks<br>Guaranteed resource discovery<br>SADQ or sub-queries<br>Applications<br>• CASA<br>• GENI<br>• P2P clouds<br>• Desktop grid<br>• File sharing | Complex & mutable resources<br>$O(\log n)$ point query cost<br>$O(\log n)$ routing state<br>Lower query cost under SADQ<br>Lower advertise cost under sub-queries<br>Latency & bandwidth match in SWORD<br>Easily extendable to support new attributes – except for LORM & SWORD<br>Privacy through anonymity<br>Track & enforce incentives & trust | $O(n)$ range query cost<br>$O(Ak)$ routing state in Mercury<br>High advertise cost if attributes change frequently – particularly under SADQ<br>High query cost under sub-queries<br>No resource match & bind<br>Index & query load imbalanced<br>Moderate resilience to churn<br>Incentives, trust, & security have to be tradeoff with anonymity<br>Lacks security unless all nodes have access to a CA |

*$A$ – No of attributes, CA – Certifying authority, $Q$ – No of queries, $R$ – No of resources, SADQ – Single-Attribute-Dominated queries

## 5 Research Opportunities and Challenges

Emerging collaborative P2P applications will require discovery and utilization of different types of resources to accomplish greater tasks that cannot be accomplished with traditional systems. Diversity in resources, application requirements, and complex inter-resource relationships present new challenges not encountered in conventional P2P systems. Our analysis of two real-world systems, PlanetLab and SETI@home, shows that multi-attribute resource and query characteristics diverge substantially from conventional assumptions [15]. For example, resource attributes and their rate of change are somewhat correlated and follow a mixture of probability distributions (e.g., Gaussian, generalized

Pareto, and generalized extreme value distributions). Furthermore, categorical attributes, e.g., CPU architecture and operating system, are highly skewed and do not fit a standard distribution. Rate of change in dynamic attributes, e.g., free CPU, memory, and bandwidth differ from one attribute/node to another, and some of the attributes changed very frequently. These factors contribute to a large resource-advertising cost. Moreover, it is typically assumed that RSs advertised by gossiping and random walk or indexed in a DHT will expire after a predetermined timeout. However, given that rate of change in dynamic attributes differ from one attribute to another and from one resource to another, it is nontrivial to determine a suitable timeout for each attribute or RS. Consistency of attribute values is important in mission critical or latency sensitive applications such as CASA and P2P clouds. Hence, to maintain the consistency of indexed resources in a DHT, each update to a dynamic attribute had to be preceded by a removal of the old attribute value. Therefore, the cost of resource advertising in DHTs effectively doubles. Such removal is not practical in unstructured P2P as an agent carrying a removal message, as a gossip or random walk, may not travel the same path taken by the previous advertisement agent. Though resources are represented using tens of attributes, they are queried using only a few attributes [15]. Dynamic attributes are highly popular where queries mostly specify attributes such as free memory, CPU load, and transmission rate. Queries are less specific as they specify only 1-7 attributes, large range of attribute values (e.g., 70% of the queries requested free disk space of 5-1000GB), and request for a large number of resources (e.g., 42% of the queries requested 50 or more resources). In practice, it is also important not to over specify a query as it limits the applicable pool of resources. We believe that these trends will remain even in collaborative P2P applications where users may not be informed enough to issue very specific queries. These findings invalidate commonly used assumptions such as independent and identically distributed attributes [45], uniform/Zipf's distribution of attribute values [22, 25], attributes having a large number of potential values [45], and queries specifying a large number of attributes and a small range of attribute values [22-23, 45]. Furthermore, simulation-based evaluation of seven resource discovery architectures using PlanetLab resource and query traces show that current designs perform poorly under real workloads [32]. Resource discovery cost of these solutions was significantly higher than predicted by prior studies that were based on conventional assumptions. Moreover, due to less specific queries, the performance bound of ring-based designs was effectively $O(n)$ instead of the well-known $O(\log n)$. Furthermore, due to highly skewed distribution of attributes and the small number of potential attribute values, resource discovery solutions (e.g., [16, 21-23, 25]) are also susceptible to significant load imbalance. Therefore, further research is needed to enhance the performance, load balancing, and robustness of resource discovery/aggregation solutions under real-workloads. It is also important to take into account the complex correlations and distributions of real-world resources and queries while designing, validating, and evaluating future resource-aggregation solutions. Therefore, traces derived from real-world systems or synthetic datasets that capture such complex relationships have to be utilized [46]. It is also useful to design solutions for the desired goals and objectives of a given collaborative application(s) as no single solution can satisfy requirements of all the possible collaborative P2P systems.

There is still a need for a cohesive solution that can select, match, and bind resources. It is desirable to combine these phases together; otherwise, optimizations in one phase could lead to conflicting results in following phases thereby reducing the opportunity for the best set of resources to collaborate. Certain applications can also compensate for lack of resources. For example, distributed data fusion in CASA can compensate for lack of bandwidth between a processing and a storage node by processing data faster (due to inherent parallelism in data fusion) to accommodate the extra delay introduced while transferring data to the storage node. It is useful to identify and support such application specific compensations within the resource aggregation solution as they can enhance the QoS and robustness. Such requirements can be represented by complex queries that are mapped to a polygon on the attribute space, e.g., $Q_2$ in Fig. 4(a). Representing and resolving such queries are not straightforward and requires tight coordination between resource selecting and matching phases.

It is challenging to capture complex inter-resource relationships accurately while introducing low overhead. Constraints such as latency and bandwidth may need to be satisfied among the selected resources as well as between the set of selected resources and the node that is trying to deploy the collaborative application. Satisfying such constraints is nontrivial in centralized, superpeer, and DHT-based solutions as the queries are resolved by third-party node(s). Latency or bandwidth measured to a third-party node is not transitive to the selected resources or to the node trying to deploy the application. However, it has been demonstrated that network coordinates [47-48], measuring performance to known landmarks [33], and random IP address sampling [49] could be used to estimate inter-resource latency without involving the third-party node. However, these measurements need to be performed a priory. Moreover, further research efforts are needed to enhance their accuracy, reduce overhead, and support other inter-resource relationships such as bandwidth, packet loss, jitter, and connectivity. Approaches such as P4P [50] and ALTO [51] will be required to infer about the physical topology and connectivity. However, these solutions have not gained much attention in production P2P systems, as users seem to be unwilling to trust ISPs' suggestions. In applications such as mobile social networks, P2P clouds, and P2P market places, awareness of peers/users social networks is also important, as it will determine what resources can be trusted, shared, with whom, and to what extent.

Solutions designed to support incentives, trust, privacy, and security in conventional P2P systems need to be extended to support interaction among multiple groups of heterogeneous/incompatible resources in collaborative P2P systems. Multi-attribute resource aggregation solutions can treat incentives and reputation values as another set of attributes. However, they are hard to enforce in unstructured P2P systems, as there are no designated nodes/resources to keep track of resources' incentives and trust values. Even if such resources can be nominated, other resources may not know how to reach them. Incentives and reputation values need to be preserved even after a resource leaves the system (due to failure or churn), as it is costly and time consuming to regain those values when the resource rejoins. Moreover, trust/reputation values may never converge if the network is prone to high churn. In [52], it has been proposed to tap into social networks of resource owners to both bootstrap the trust network and during periods of high churn. However, alternative approaches are needed for networks that do not have access to social networks. [42] is an initial step in this direction; however, its efficiency and reliability need to be enhanced. Furthermore, security, integrity, and accountability of nodes maintaining incentives and trust values are of utmost importance, as they can become easy targets for attacks. Guidelines need to be identified for determining credits/payments and reputation scores for heterogeneous resources. For example, while both a radar and a set of rain gauges are important for weather monitoring, formally evaluating the cost and perceived value of such systems in a consistent manner is difficult. Moreover, while a computed result must be always accurate, accuracy of sensor data is dependent on many dynamic parameters. Formal analysis of incentive schemes such as [36] need to be extended to understand under what conditions a collaborative P2P system will be robust and when will it collapse. Trust can be task based or situational, where two resources that would not otherwise collaborate with each other may collaborate for a specific cause, e.g., a mobile social network deployed after a disaster. Hence, it is important to support both long-term trust values and task-based ones. Centralized or superpeer-based index is not desirable when anonymity is important. Anonymity is in conflict with incentives, trust, and security; hence, it is important to look for distributed solutions that overcome this limitation. Existing solutions rely mostly on digital certificates (self-signed, signed by a CA, or chain of CAs) to enforce access control, integrity, and other security issues. However, it is nontrivial to enforce those unless at least a subset of the resources has access to a CA(s). Moreover, access rights should be transferable as autonomous agents may be deployed to carry out collaborative applications on behalf of their users (e.g., a sensor directly talking to a processing node). Therefore, solutions that are specifically designed for isolated, collaborative, and distributed environments are needed. Because the key phases of resource collaboration and incentives, trust, privacy and security are essential elements of a collaborative P2P system, their performance should also be evaluated in the context of the overall system. Issues related to incentives, trust, privacy, and security might seem to be overweighting the benefits of collaborative P2P systems. However, with the right tools and incentives in place, it will be more useful, efficient, and rewarding to accomplish a greater task through collaboration.

Distributed resource binding could create conflicts among peers hence there is also a need for distributed resource negotiation. Heterogeneous P2P systems have diverse resources that are characterized by tens of attributes (e.g., CASA and GENI). Such a large number of attributes hinders the performance of all the existing solutions. Moreover, our findings show that 80% of the queries in PlanetLab specify only two attributes [32]. Therefore, while indexing resources, it is sensible to depend on a small set of attributes that are essential to characterize a resource or use dimension reduction techniques. Rest of the attributes, including some of the dynamic ones, could be checked at individual peers, which is anyway necessary if resource binding is required (except in centralized and superpeer architectures). However, performance and QoS of latency sensitive applications such CASA and P2P clouds depend on dynamic attributes; hence, a primary subset of dynamic attributes need to be advertised at a higher granularity.

Centralized solutions are becoming more feasible, affordable, and reliable due to the recent advancement in distributed datacenter technologies. When applicable/feasible, centralized solution is a better option, as it can advertise, select, match, and bind resources, and enforce incentives, trust, and security. However, it lacks privacy and prone to attacks, as the centralized node is the only source of trust in the system. Furthermore, with increased levels of integration, even the systems within datacenters exhibit attributes of distributed systems. Superpeers provide probably the best design choice as they are distributed and simultaneously support resource advertising, selecting, matching, and binding [15] yet further research is needed to better support resource matching, inter-superpeer resource sharing, and to obtain predictable performance. Single-attribute-dominated queries on a single ring [16, 22-23] is suitable for structured P2P systems. However, issues related to skewed distribution of attributes, high cost of advertising dynamic attributes, load balancing, matching, and binding need to be addressed. Moreover, standard approaches such as caching and replication are not that effective as attribute values are highly dynamic. Hence, alternative solutions that can benefit from the skewed distribution of resources and queries need to be developed. Addressing all these challenges is essential to realize the full potential and capabilities of collaborative P2P systems.

# 6   Summary

Collaborative P2P systems require the ability to aggregate complex groups of heterogeneous, multi-attribute, distributed, and dynamic resources as and when needed. The process of resource aggregation is illustrated by identifying a set of key phases and their relevance to emerging collaborative P2P applications exemplified by systems such as CASA, GENI, P2P clouds, and mobile social networks. State-of-the-art resource discovery/aggregation architectures were compared with respect to their ability to support key phases of resource aggregation, lookup overhead, load balancing, scalability, incentives, trust, privacy, and security. It was observed that existing architectures are applicable only under specific conditions, cannot cohesively support resource selection, match, and bind, and perform poorly under realistic workloads. These limitations points out the need for novel solution(s) that can effectively aggregate resources in real-world collaborative P2P systems. We are currently developing a multi-attribute resource aggregation solution that supports all the key phases of resource aggregation while overcoming limitations in existing architectures.

# Acknowledgements

This research is supported in part by the Engineering Research Center program of the National Science Foundation under NSF award number 0313747.